\documentclass[draft, english]{volcanica-template} 

\newcommand{\Title}{Spectral stability of V2 centres in sub-micron 4H-SiC membranes} 
\newcommand{\shortTitle}{Spectral stability of V2 centres in sub-micron 4H-SiC membranes} 
\newcommand{\Author}{J. Heiler et al.\xspace} 
\author[{{\affiliation{1},\affiliation{2},\affiliation{3}}}]
{Jonah Heiler~\orcidaffil{0009-0000-4621-4782}\SharedAuthor}

\author[{{\affiliation{1}}}]
{Jonathan K\"orber~\orcidaffil{0000-0002-7531-0295}\samethanks
\Email{jonathan.koerber@pi3.uni-stuttgart.de}} 

\author[{{\affiliation{1}}}]
{Erik Hesselmeier~\orcidaffil{0000-0002-4560-1745}\samesamethanks}

\author[{{\affiliation{1}}}]
{Pierre Kuna}

\author[{{\affiliation{1}}}]
{Rainer St\"ohr~\orcidaffil{0000-0002-3476-344X}}

\author[{{\affiliation{4}}}]
{Philipp Fuchs~\orcidaffil{0000-0002-9966-6093}}

\author[{{\affiliation{5}}}]
{Misagh Ghezellou~\orcidaffil{0000-0002-1969-3324}}

\author[{{\affiliation{5}}}]
{Jawad Ul-Hassan~\orcidaffil{0000-0001-9537-2226}}

\author[{{\affiliation{6}}}]
{Wolfgang Knolle}

\author[{{\affiliation{4}}}]
{Christoph Becher~\orcidaffil{0000-0003-4645-6882}}

\author[{{\affiliation{1},\affiliation{2},\affiliation{3}}}]
{Florian Kaiser~\orcidaffil{0000-0002-5844-1779}
\Email{florian.kaiser@list.lu}}

\author[{{\affiliation{1},\affiliation{7}}}]
{Jörg Wrachtrup}

\affil[{{\affiliation{1}}}]{					
3rd Institute of Physics, University of Stuttgart, 70569 Stuttgart, Germany.
}

\affil[{{\affiliation{2}}}]{					
Materials Research and Technology (MRT) Department, Luxembourg Institute of Science and Technology (LIST), 4422 Belvaux, Luxembourg.}

\affil[{{\affiliation{3}}}]{					
Department of Physics and Materials Science, University of Luxembourg, 4422 Belvaux, Luxembourg}

\affil[{{\affiliation{4}}}]{					
Universität des Saarlandes, Fachrichtung Physik, Campus E2.6, 66123 Saarbrücken, Germany.
}
\affil[{{\affiliation{5}}}]{					
Department of Physics, Chemistry and Biology, Linköping University, 581 83 Linköping, Sweden.
}
\affil[{{\affiliation{6}}}]{					
Leibniz-Institute of Surface Engineering (IOM), Leipzig, Germany.}

\affil[{{\affiliation{7}}}]{                    
Max Planck Institute for Solid State Research, Stuttgart, Germany.
}

\begin{document}


\FrontMatter{\protect{
\noindent Colour centres in silicon carbide emerge as a promising semiconductor quantum technology platform with excellent spin-optical coherences.
However, recent efforts towards maximising the photonic efficiency via integration into nanophotonic structures proved to be challenging due to reduced spectral stabilities.
Here, we provide a large-scale systematic investigation on silicon vacancy centres in thin silicon carbide membranes with thicknesses down to \SI{0.25}{\micro\meter}.
Our membrane fabrication process involves a combination of chemical mechanical polishing, reactive ion etching, and subsequent annealing.
This leads to highly reproducible membranes with roughness values of \SIrange{3}{4}{\angstrom}, as well as negligible surface fluorescence.
We find that silicon vacancy centres show close-to lifetime limited optical linewidths with almost no signs of spectral wandering down to membrane thicknesses of \SI{\sim 0.7}{\micro\meter}.
For silicon vacancy centres in thinner membranes down to \SI{0.25}{\micro\meter}, we observe spectral wandering, however, optical linewidths remain below \SI{200}{\mega\hertz}, which is compatible with spin-selective excitation schemes.
Our work clearly shows that silicon vacancy centres can be integrated into sub-micron silicon carbide membranes, which opens the avenue towards obtaining the necessary improvements in photon extraction efficiency based on nanophotonic structuring.
}}[]{}

\section*{INTRODUCTION}\label{sec:introduction}
Optically active spins in solids have implemented quantum technology applications across all fields~\cite{Awschalom2018}.
Following up on landmark experiments with nitrogen-vacancy (NV) centres in diamond, a current technology trend focuses on colour centres in industrial semiconductor platforms, to benefit from advanced (nano-)manufacturing at the wafer-scale.
Recent work identified colour centres and optically active dopants across a variety of semiconductor materials, such as GaN~\cite{BishopGaN2022,LuoGaN2023}, AlN~\cite{CannonAlN2023}, ZnO~\cite{ChoiZnO2015,NiaourisZnO2022}, Si~\cite{DurandSi2021,HigginbottomSi2022,HollenbachSi2022}, and SiC~\cite{Castelletto2020,Castelletto2021}.
For the latter two, nuclear spin-free crystals based on isotope purification are already established, which resulted in multi-second colour centre spin coherence times~\cite{Anderson5s2022}.
Additionally, colour centres in SiC have demonstrated nearly lifetime limited optical transitions, which enabled high-fidelity spin-photon interfaces that are suitable for quantum communication~\cite{Morioka2020}.

A current issue for quantum applications with individual colour centres concerns the efficient collection of fluorescence photons: The high refractive index of the above-mentioned materials ($n>2$, and even $n>3.5$ for Si) leads to total internal reflection at the interface crystal-air.

Promising advances towards improved light collection efficiency were recently demonstrated with negatively charged silicon vacancy ($\rm V_{Si}^-$) centres in 4H-SiC, based on nanostructured solid-immersion lenses~\cite{Nagy2019}, nano-pillars~\cite{Radulaski2017}, waveguides~\cite{Babin2022}, photonic crystal cavities~\cite{Lukin2020}, and disk resonators~\cite{Lukin2023}.
In most studies, nearly lifetime limited optical linewidths were observed, however, with a tendency for slow spectral wandering in nanoscopic devices.
This currently prevents setting up networks of spectrally orchestrated $\rm V_{Si}^-$ centres in SiC.

In this work, we show a systematic investigation of $\rm V_{Si}^-$ centres in SiC membranes with several thicknesses ranging from \SI{0.13}{\micro\meter} to \SI{2.36}{\micro\meter}.
Our study reveals almost no degradation of optical coherences down to membrane thicknesses of \SI{0.7}{\micro\meter}, which is compatible with emitter integration into resonant enhancement structures, e.g., planar optical antennas~\cite{Fuchs2021} or external Fabry-Pérot cavities~\cite{Riedel2017,Haeussler2019,Salz2020,Ruf2021}.
Such structures could permit to efficiently orchestrate high-quality photonic interference between multiple emitters~\cite{HermansTele2022}.
In membranes below \SI{0.7}{\micro\meter} thickness, we observe a tendency for slow spectral wandering.
Also, at a \SI{0.25}{\micro\meter} membrane thickness, we measure a slight linewidth broadening to \SI{\sim 200}{\mega\hertz}, which, however, is still compatible with spin-selective excitation.
Thus, $\rm V_{Si}^-$ centres in SiC membranes could be used for high-fidelity electron-nuclear-spin-related experiments, such as quantum registers~\cite{Bradley2019}.

\section*{RESULTS AND DISCUSSION}\label{sec:results}
\subsection*{Membrane fabrication}
\begin{figure*}[tb]
\includegraphics[width=\textwidth]{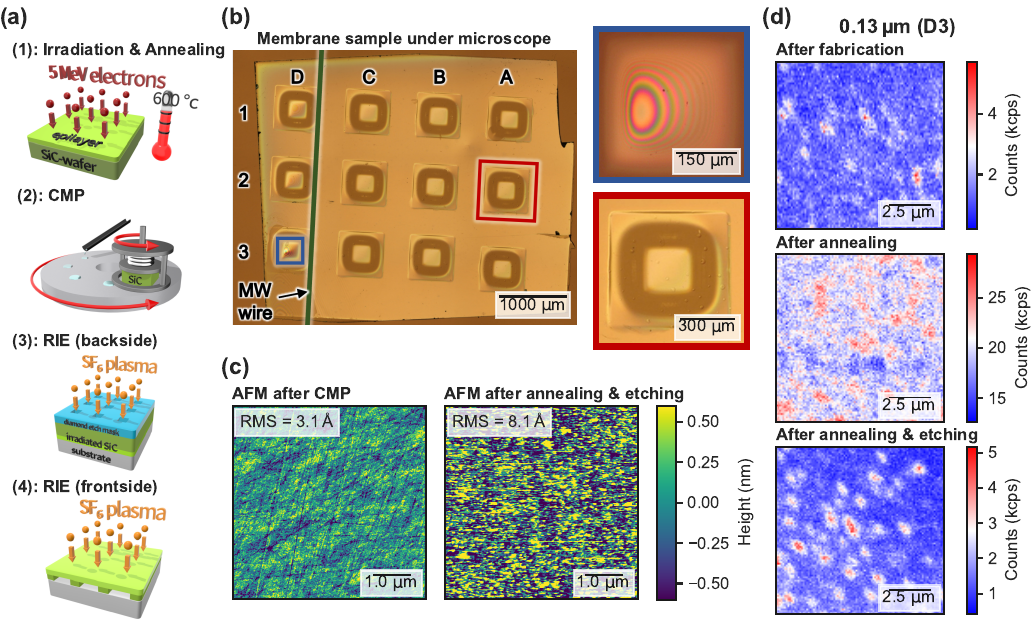}
\centering
\caption[]{\textbf{Membrane fabrication scheme and surface characterisation.} \textbf{(a)} Membrane fabrication: (1) $\rm V_{Si}^-$-creation via irradiation with electrons at an energy of \SI{5}{\mega\electronvolt} and a dose of \SI{2}{\kilo\gray} and subsequent annealing at \SI{600}{\degreeCelsius}.
(2) Lapping and CMP of the wafer down to a total sample thickness of around \SI{40}{\micro\meter}.
(3) Fabrication of \SI{35}{\micro\meter} trenches using RIE with an SF$_\mathrm{6}$-plasma.
The trenches are defined by a diamond hard mask with square openings (blue).
(4) Final thinning of the membranes after flipping the sample with the same RIE process already used in (3). 
\textbf{(b)} Light microscope images of the sample after fabrication.
The inset with a blue frame shows a zoom of the thinnest membrane D3, where visual interference fringes indicate a thickness gradient.
The red inset shows a zoom of membrane A2 with visible defects on the surface that appeared after an SF$_\mathrm{6}$ soft-ICP etching that was done as part of the post fabrication.
\textbf{(c)} Surface topography on an arbitrary position of our sample after the final polishing of step (2) (left panel) and on membrane D3 (right panel) after the post fabrication, measured with a commercial AFM.
The extracted RMS-roughness are shown on the top left of each panel.
\textbf{(d)} Photoluminescence scan under \SI{785}{\nano\meter} excitation at \SI{0.4}{\milli\watt} on an arbitrary position of membrane D3 after the main fabrication steps (upper), the post fabrication annealing (centre) and an additional soft-ICP etching step (lower).}
\label{Fig1}
\end{figure*}
The steps that are used to fabricate membranes with integrated V2 centres are
inspired by previous works \cite{Lukin2020,Heupel2020,Fuchs2021}, and are
sketched in Figure \ref{Fig1}(a).
Our cleaning procedures in between steps are described in Supplementary Note 1.
Similar to previous studies \cite{Nagy2019, Babin2022}, we use electron irradiation followed by thermal annealing to create single $\rm V_{Si}^-$ centres.
We note that we decided to create V2 centres before the main fabrication steps by purpose to study the influence of our fabrication on the optical properties of colour centres.

Consecutively, we remove material from the wafer side (opposite to the high-quality epilayer) by lapping until we reach a total sample thickness of \SI{\sim50}{\micro\meter}.
To remove scratches and subsurface damage induced by the lapping, a series of chemical-mechanical polishing (CMP) steps is used, yielding a total thickness of \SI{\sim42}{\micro\meter}.
Our polishing yields a root mean square (RMS) surface roughness of \SIrange{3}{4}{\angstrom}, shown as a representative standard in the left panel of Figure \ref{Fig1}(c) and measured with an atomic force microscope (AFM) on an area of $5 \times \SI{5}{\micro\meter}^2$. 

To create (sub-)µm-thin membranes, we use reactive-ion etching (RIE) with an SF$_6$-plasma at an etch-rate of \SIrange{56}{60}{\nano\meter\per\minute}.
At first, we place a diamond hard mask with 12 square openings of \SI{350}{\micro\meter} length, see Supplementary Note 2 for details, on the wafer side of the sample and create trenches with a depth of \SI{\sim35}{\micro\meter}.
By using the diamond mask we retain outer frames at a thickness of \SI{\sim42}{\micro\meter} for mechanical stability and easier handling at subsequent steps of the sample, while the 12 open regions serve as thin membranes.
Consecutively, the diamond mask is removed and the sample is flipped, followed by a series of short etching steps for the final thinning.
To retain control over the membrane thickness at the final etching steps, we measure the thickness minimum of each membrane, labelled as shown in Figure \ref{Fig1}(c), with a home-built white-light-interferometer (WLI, details in Supplementary Note 3).

After the main fabrication and optical characterisation, that will be discussed in the next section, we perform a second annealing of our sample with the same parameters as the first annealing.
We perform photoluminescence (PL) scans of our sample at room temperature with an excitation laser of \SI{728}{\nano\meter} and \SIrange{350}{450}{\micro\watt} before the objective.
We find that the background fluorescence of our sample increased by one order of magnitude after the annealing, which can be seen by comparing the upper and centre panel of Figure \ref{Fig1}(d).
Furthermore, we observe very slow bleaching of the parts of the surface that were exposed to the \SI{728}{\nano\meter} laser.
To clean the surface of the parasitic fluorescence, we perform a short soft-ICP etching step on the growth side of the epilayer with an SF$_6$-plasma. We removed \SI{\sim 30}{\nano\meter} of material, which resulted in an only slightly increased background fluorescence level compared to the one after the main fabrication, as shown in the lower panel of Figure \ref{Fig1}(d) and Supplementary Figure 3.
However, as displayed in the right panel of Figure \ref{Fig1}(c), this soft-ICP etching induced a grain-like surface with an increase of the surface roughness by a factor of \SI{\sim2.5}{}.
We suspect that the annealing step created contamination on the surface which was partially imprinted on the SiC surface at our soft-ICP step.
We emphasise that the induced roughness is not related to the RIE of our main fabrication in all likelihood, since we know from previous samples that our process yields a surface roughness comparable to the one after polishing (Supplementary Notes 5 and 6 for details).
After all fabrication steps, the thinnest membrane of our sample exhibits a minimum thickness of \SI{\sim0.13}{\micro\meter} and the thickest membrane a minimum thickness of \SI{\sim2.36}{\micro\meter}.
We attribute this inhomogeneity to our lapping and polishing steps.
For simplicity, we now always refer to the membrane thicknesses after the soft-ICP step.

\subsection*{Emission spectra in membranes and bulk}

\begin{figure*}
\includegraphics[width=\textwidth]{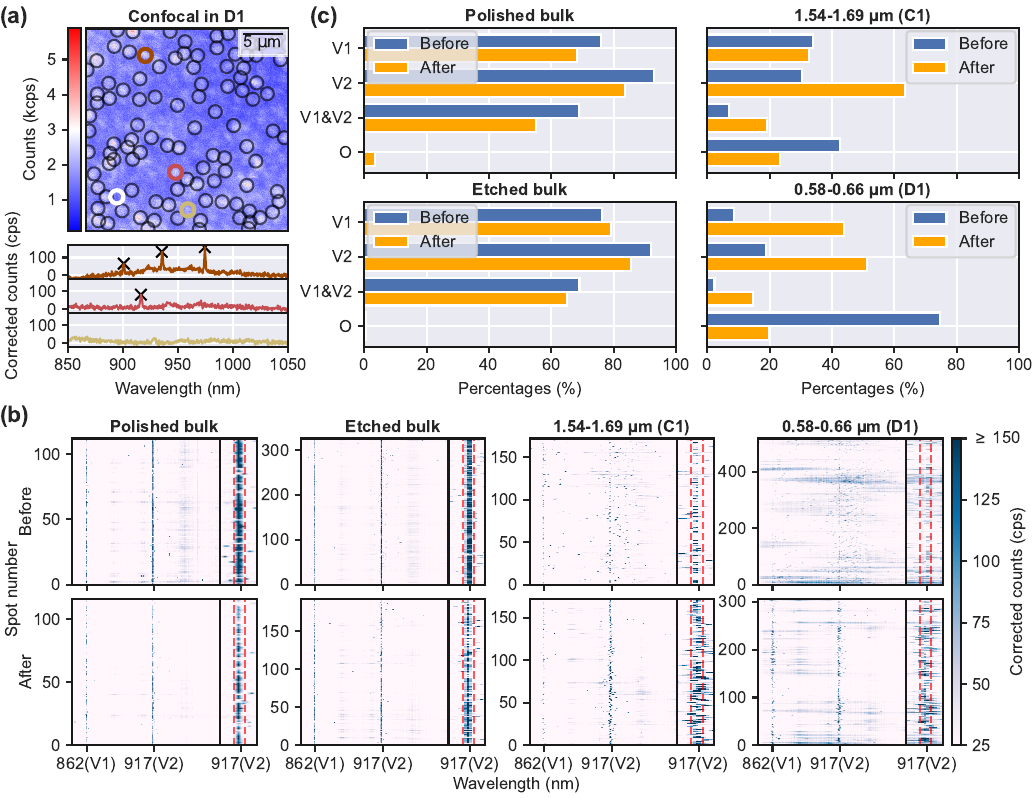}
\centering
\caption[]{
\textbf{Spectra acquisition, statistics, and comparison for bulk and membrane material.}
\textbf{(a)} Confocal map of a $25 \times \SI{25}{\micro\meter\squared}$ region in membrane D1 with circles marking the manually selected positions of recorded photoluminescence spectra.
The bold white circle indicates where the background correction spectrum was recorded, which we subtracted from all subsequent measurements. The brown, red, and yellow circles correspond to three representative spectra, which are shown below: brown/top: multiple peaks within one spectrum; red/middle: a single V2-peak; yellow/bottom: no peak.
\textbf{(b)} Stacked photoluminescence spectra from different spots with peaks on a bulk sample that was only polished, on the bulk part of the etched membrane sample, on the membrane C1, and on the membrane D1. 
The upper (lower) panels show data before (after) the post-fabrication processing.
Each line corresponds to a background corrected spectrum taken at one spot in the respective region.
The insets on the right zoom into the wavelength interval \SIrange{914}{919}{\nano\meter} where dashed red lines indicate the region \SIrange{915.7}{917.1}{\nano\meter} in which peaks are ascribed to emission from a V2 centre for (c). The respective membrane thickness is given in the title.
Peaks in the range of \SIrange{861.0}{862.4}{\nano\meter} are considered to be V1 centres, the reasoning for these intervals is given in the main text.
\textbf{(c)} Percentages of V1-peaks, V2-peaks, both V1- and V2-peaks (V1\&V2), and unidentified peaks (O) in the spectra, obtained as in (a), before and after the post-fabrication processing.
}
\label{Fig2}
\end{figure*}

The emission spectrum of V1 (V2) centres in 4H-SiC at cryogenic temperatures is characterised by a sharp zero-phonon line (ZPL) at \SI{862}{\nano\meter} (\SI{917}{\nano\meter}) and a broad phonon side band (PSB) \cite{Wagner2000, Ivady2017, Udvarhelyi2020}. 
To investigate the influence of our fabrication process on the implanted silicon vacancy centres, we record over 100 emission spectra in each of the regions. These regions comprise the only-polished bulk, the etched bulk, and the four membrane windows C1, D1, D2, and D3 with respective thicknesses of \SIrange{1.54}{1.66}{\micro\meter}, \SIrange{0.58}{0.63}{\micro\meter}, \SIrange{0.63}{0.72}{\micro\meter}, and \SIrange{0.13}{0.25}{\micro\meter}. Thicknesses were measured by WLI about \SI{50}{\micro\meter} next to the respective membrane centres (see 
Supplementary Note 7), which corresponds to the areas in which we performed the investigations on the colour centres.
Additional data for the latter two membranes can be found in Supplementary Note 8.
For acquisition of spectra we developed an automatised software that optimises the focus and records a spectrum on each manually selected point of a high resolution confocal map, as shown in Figure \ref{Fig2}(a).

Next, we extract the spectral positions of peaks in these spectra and notice that most peak positions in only polished bulk and etched bulk material align at the ZPL of the silicon vacancy centres, while the spectral positions of peaks in membranes are wide-spread over the detection window.
This difference is visualised for membranes C1 and D1 in Figure \ref{Fig2}(b).
Stacking spectra with at least one peak in a two-dimensional plot results in two clear lines for V1- and V2-ZPL in the bulk case.
For the membranes, however, those lines are hardly visible in this representation, indicating the fabrication process induced crystal damage in vicinity to the silicon vacancies that altered their structure and therefore their ZPL emission frequency.
Motivated by this insight, we decide to continue with an annealing step (same as for the color center creation, see methods) in effort to heal the crystal and recover the altered silicon vacancy centres.
Note that we only anneal the sample with membranes, while keeping the polished bulk as a reference. However, we expose both samples to a subsequent short SF$_6$ ICP etching step.
The stacked spectra after those post-fabrication steps are depicted in the bottom row of Figure \ref{Fig2}(b). As expected, in the bulk sample, there are essentially no observable changes. Yet, the emission spectra in the annealed membrane sample now show a significantly increased number of emission lines aligning around the typical V1- and V2-ZPL.

To quantify this change, we assign each peak in the spectra to a V1 (V2) centre if its spectral position lies maximum two spectrometer pixels, i.e. \SI{0.7}{\nano\meter}, away from the position of the V1- (V2-) ZPL in bulk material.
We choose this cut-off, since we measured photoluminescence excitation (PLE) on V2 centres with ZPL peaks at most two pixel away from the bulk position.
The interval for V2 centres is indicated by dashed read lines in the insets of Figure \ref{Fig2}(b).
Every peak out of both intervals is assigned to the category "other defects" (O) in the overall statistics, which is shown in Figure \ref{Fig2}(c).
This statistics also includes an extra category for simultaneous appearances of V1- and V2-peaks within a diffraction limited spot, which is a rather regular observation that we made.

Overall, the percentage occurrences show, as expected, that bulk material has the highest amount of $\rm V_{Si}^-$ centres at the V1- and V2-position.
For membranes, after the first fabrication step, up to \SI{74}{\percent} less V1- and V2-peaks are observed, and many new emission peaks appear at different wavelengths. As a general trend, we find that the relative amount of $\rm V_{Si}^-$ centres decreases with reduced membrane thickness.
After post-fabrication (annealing, short ICP), the amount of $\rm V_{Si}^-$ centres in the membranes increases up to a factor of five, indicating that some crystal damage from the fabrication was repaired.
Even though the spots investigated before and after post fabrication are not exactly the same, the high number of investigated spots enables us to identify this trend.
To investigate the spectral features of V2 centres in more detail, we perform additional PLE measurements.

\subsection*{Resonant laser excitation measurements in membranes and bulk}
\begin{figure*}
\includegraphics[width=\textwidth]{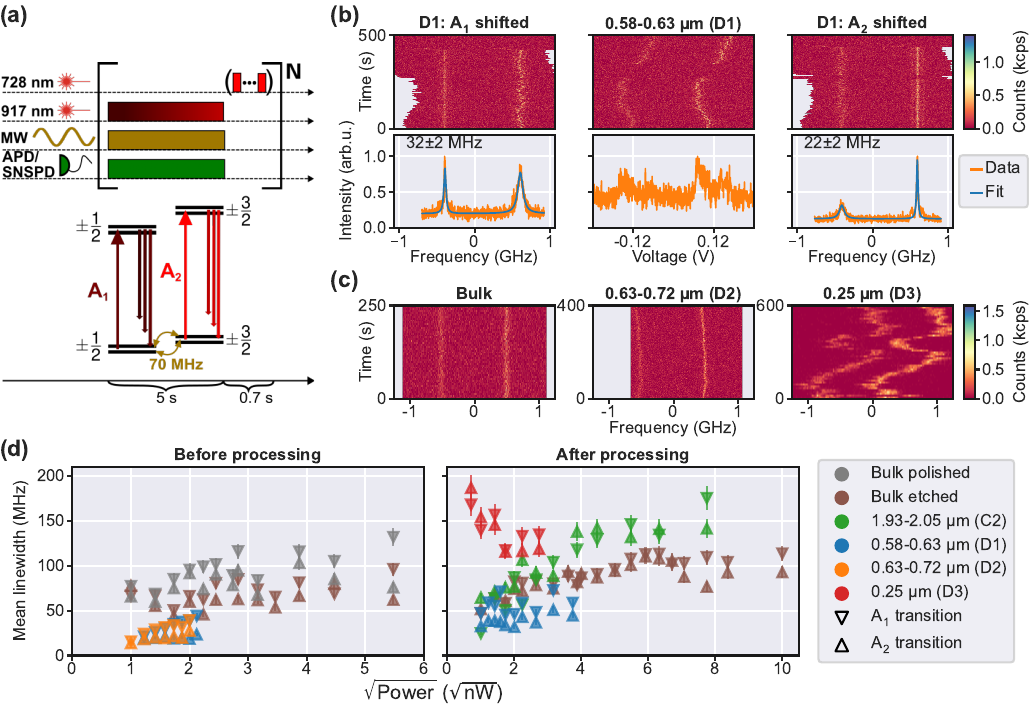}
\centering
\caption[]{
\textbf{PLE measurement scheme, PLE examples with linewidth evaluation, and power dependent linewidths for membranes as well as bulk material.}
\textbf{(a)} Spin-3/2 level scheme of the $\rm V_{Si}^-$ centre together with the measurement sequence of the PLE scan in (b) consisting of $N=100$ single line scans of \SI{5}{\second} each with an integration time of \SI{\sim5}{\milli\second} per point (see methods for details).
\textbf{(b)} Exemplary linewidth extraction from a PLE scan of one emitter in membrane D1 at a resonant laser power of \SI{2}{\nano\watt}.
The middle panels show the scanned lines recorded over time and the normalised sum of all recorded lines over the applied voltage.
The left (right) panels show the A$_1$ (A$_2$) shifted scanned lines and their sum over the frequency, fitted with a Lorentzian profile to extract the A$_1$ (A$_2$) linewidth.
\textbf{(c)} PLE scans at \SI{2}{\nano\watt} resonant excitation power for single emitters in the etched bulk, membrane D2, and membrane D3.
The abscissa shows the frequency obtained from a conversion analogous to (b) and the respective membrane thicknesses are given in the title.
\textbf{(d)} Mean A$_1$ and A$_2$ linewidths, extracted as in (b), of emitters before and after post-fabrication processing in only polished bulk, etched bulk, as well as membranes C2, D1, D2, and D3 as a function of the root of the excitation power.
The error bars correspond to a standard deviation of the splitting of \SI{75}{\mega\hertz} (see methods section).
}
\label{Fig3}
\end{figure*}

As depicted in Figure \ref{Fig3}(a), we record the PLE spectra by sweeping a narrowband tunable laser across the optical transitions within the V2-peak, while detecting the PSB emission.
In addition, we apply continuous microwaves (MW) at \SI{70}{\mega\hertz} to mix the spin populations in the ground states, which counteracts spin-pumping processes. The MW signal is provided through a \SI{50}{\micro\meter} copper wire spanned across our sample nearby the membranes D1, D2, and D3, shown in Figure \ref{Fig1}(b). 
We repeat the PLE scheme $N=$ \SIrange{20}{100}{} times for each emitter and different laser powers, to observe its stability over time and to get more statistics for the linewidth extraction.
Four exemplary PLE scans at \SI{2}{\nano\watt} excitation power in etched bulk, membrane D2, D3, and D1 are depicted in Figures \ref{Fig3}(b) and \ref{Fig3}(c).
The scan in D3 shows a significantly better signal-to-noise ratio (SNR) than the other three examples since it was recorded after post-fabrication and we upgraded our detector from an avalanche photo-diode (APD) to a superconducting nanowire single-photon detector (SNSPD) during this processing.
Furthermore, the SNR reduces with smaller excitation powers, which challenges an accurate determination of the linewidths from single line scans.
To obtain the linewidth from PLE scans at low SNR conditions, we need to average data over multiple scans. To additionally account for spectral wandering, we align the positions of all A$_1$ (A$_2$) peaks based on single line fits and then perform another Lorentzian fit on the sum of all single lines.
From the latter fit, we extract the A$_1$ (A$_2$) linewidth as the full width at half maximum FWHM by converting the x-axis to frequencies based on the \SI{1}{\giga\hertz} separation of the A$_1$ and A$_2$ transition \cite{Banks2019, Tarasenko2018}.
An example of this process is shown in Figure \ref{Fig3}(b).
More details can be found in the methods section.

To be consistent with the linewidth extraction, we use this procedure for every PLE scan.
Figure \ref{Fig3}(d) shows averaged linewidths for different regions as a function of the resonant laser power.
An overview of all extracted linewidths for individual emitters is displayed in Supplementary Note 9.
The left-sided panel of Figure \ref{Fig3}(d) shows mean linewidths for 14 emitters before the post-processing steps (annealing, short ICP), comprising 3 $\rm V_{Si}^-$ centres in the polished bulk, 2 in the etched bulk, 4 $\rm V_{Si}^-$ centres in membrane D1, and 5 in D2.
The right-sided panel shows mean linewidths after the final processing step for 3 $\rm V_{Si}^-$ centres in the etched bulk and 2 in membrane D1, as well as the linewidths of 1 emitter measured in membranes C2 and D3, respectively.
Although post-processing did approximately triple the amount of V2-peaks across all membranes, we found that measuring power-dependent linewidths became more difficult due to increased spectral jumping.
I.e., before post-processing, we found that only $4/18 \sim \SI{22}{\percent}$ of the V2 centres, for which we could observe a PLE signal, showed spectral jumps.
However, after post-processing, $22/29 \sim \SI{76}{\percent}$ of such V2 centres showed spectral jumping, even at low resonant laser powers (few nW).
An explanation could be that annealing repaired the lattice enough to recover the original ZPL wavelengths for several V2 centres, which we have not considered before.
However, these recovered V2 centres are still surrounded by some remaining lattice damage that gives rise to charge fluctuations and subsequent wavelength jumps.
Since emitters in membranes are close to surfaces, it is indeed unlikely that a simple annealing step can recover a pristine quality.

The mean linewidths in the membranes D1 and D2 before the post-fabrication showed the expected tendency to slightly increase with power up to about \SI{50}{\mega\hertz} at a power of \SI{4.5}{\nano\watt}.
They are, without exception, narrower than the mean linewidths obtained in the bulk material at the same power that go up to \SI{132}{\mega\hertz} at greater powers.
For a few data sets recorded at the lowest laser power, we obtain fitted linewidths slightly below the lifetime-limited values of \SI{26}{\mega\hertz} (\SI{14}{\mega\hertz}), calculated as $(2\pi\tau)^{-1}$ from the A$_1$ (A$_2$) lifetimes $\tau=\SI{6.1}{\nano\second} \, (\SI{11.3}{\nano\second})$ \cite{DiPaper}.
This phenomenon has been studied in detail in a previous work \cite{Orphal2023}, where it was shown that measurements at low signal-to-noise can lead to an underestimation of the resonant linewidths.
This is also corroborated by our studies.
After the post-fabrication step, we upgraded our detectors to SNSPDs, which improved the signal-to-noise significantly.
No sub-lifetime limited linewidths were recorded anymore, with the linewidths in the membrane D1 going down to \SI{29}{\mega\hertz} (see Supplementary Note 9).
The linewidths in bulk material show no significant change and are similar to the linewidths in the \SIrange{1.93}{2.04}{\micro\meter} membrane C2.
In the membrane D3 at \SI{0.25}{\micro\meter} thickness, we measure slightly broader linewidths in the range of \SIrange{116}{187}{\mega\hertz}.

Overall, for all power series measurements, we find little or no increase of the linewidth with the optical power.
This indicates that the laser powers are not yet in a regime of significant power broadening.
Especially in sub-micron membranes, we observed defects to ionise more rapidly at higher laser powers, which limited our measurements to powers of $P\lesssim \SI{10}{\nano\watt}$.
An explanation could be a less stable charge environment for emitters in surface proximity due to an increased abundance of charge traps and/or band bending.
Additionally, our data for the emitter in the thinnest membrane D3 (red triangles) shows a parasitic trend of decreased linewidths at higher excitation power.
As we detail in Supplementary Note 10, this observation is explained by an increased rate of spectral jumps and ionisations at higher powers.
This leads to asymmetric line shapes for which the Lorentzian fit function systematically underestimates the linewidth.

\begin{figure}
    \includegraphics[width=\columnwidth]{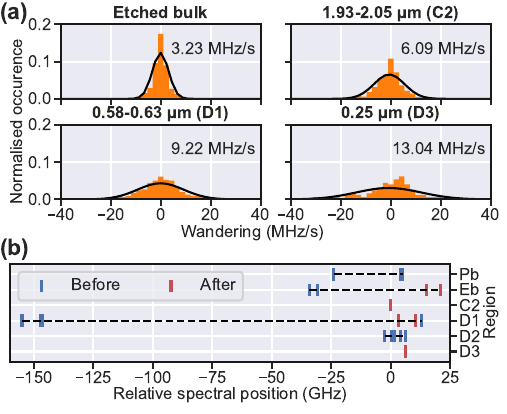}
\caption[]{
\textbf{Wandering histogram and inhomogeneous distribution of the PLE positions.}
\textbf{(a)} Histograms of the the spectral wandering in consecutive PLE lines of emitters in etched bulk material and membranes C2, D1, and D3 after the post-fabrication annealing and soft-ICP etching.
The black curve represents a Gaussian distribution with the standard deviation and mean value of the wandering data.
\textbf{(b)} Relative laser starting position for the performed PLE scans with respect to the region before and after the post-fabrication processing.
The relative position of zero corresponds to an absolute frequency of \SI{327.100}{\tera\hertz}.
Pb and Eb stand for polished bulk and etched bulk, respectively.
}
\label{Fig4}
\end{figure}

We also find that emitters in thinner membranes show an increased tendency for (slow) spectral wandering, which can be explained by the $\rm V_{Si}^-$ centre's sensitivity to electric fields and charge noise \cite{Lukin2020Stark, Ruehl2020}.
Qualitatively, this can be seen in Figures \ref{Fig3}(b) and \ref{Fig3}(c), where lines in the bulk material show no visible spectral wandering. Aligned with the membrane thickness, PLE lines show an increased wandering from membrane D2 (\SIrange{0.63}{0.72}{\micro\meter}), to D1 (\SIrange{0.58}{0.63}{\micro\meter}), to D3 (\SIrange{0.13}{0.25}{\micro\meter}).
We quantify the wandering by calculating the peak position change per time in two consecutive PLE line scans.
The evaluated data is depicted in Figure \ref{Fig4}(a) in form of histograms.
From the standard deviation of each data set, we see a clear correlation between an increased wandering and thinner structures, which we expect due to the closer proximity to the surface.

Lastly, we investigate the inhomogeneous distribution of our spectral PLE positions.
For this purpose, we show the laser frequencies, from where we started the PLE power series in Figure \ref{Fig4}(b).
We see a distribution of \SI{176}{\giga\hertz} in between all regions, with a largest contribution coming from the membrane D1 with \SI{168}{\giga\hertz}.
This spread can be attributed to a combination of the fabrication process and strain in the membranes and matches our observation from the emission spectra.
Additionally, we observe a distribution of \SI{29}{\giga\hertz} for emitters measured in the only polished bulk material, which is comparable to previous reports \cite{Babin2022}, thus confirming that deep bulk emitters are not influenced by the fabrication. This is in line with the observation that sub-surface damage is only expected within the first \SIrange{2}{5}{\micro\meter}, even before polishing \cite{Geng2022, Wang2023}.

In conclusion, we developed a reproducible method for fabrication of sub-micron 4H-SiC membranes with low surface roughness and integrated V$_\mathrm{Si}^{-}$ centres.
As reported previously \cite{Babin2022}, we also observe a decrease in the emitter density after the fabrication of sub-micron structures.
Additionally, we showed that spectral shifts of the ZPLs induced by our initial fabrication steps can be recovered by post-processing annealing.
However, the annealing process led to a reduced percentage of V$_\mathrm{Si}^{-}$ centres with stable PLE lines, which may require further studies on optimal post-fabrication processes. These should potentially include surface treatments to recover a pristine SiC surface quality.

All of our investigated emitters with stable PLE showed narrow linewidths below \SI{200}{\mega\hertz}, which is compatible with spin-selective excitation schemes. Moreover, all emitters in the \SI{\sim0.65}{\micro\meter} thick membrane showed linewidths below \SI{100}{\mega\hertz}, which is comparable with recent studies on V$_\mathrm{Si}^{-}$ centres in 4H-SiC integrated in triangular waveguides with a height of \SI{\sim500}{\nano\meter} and a width of \SI{\sim1000}{\nano\meter} \cite{Babin2022}, as well as \SI{\sim500}{\nano\meter} thick disk resonators \cite{Lukin2023}.
At the lowest powers, our measured linewidths in the \SIrange{0.6}{2}{\micro\meter} thick membranes are at most $4.6$ times larger than the lifetime limit (\SI{65}{\mega\hertz} to \SI{14}{\mega\hertz}).
For comparison, the linewidths of NV centres in a \SI{3.8}{\micro\meter} diamond membrane, have in average a 6.6 times broadened linewidth (\SI{86}{\mega\hertz} to \SI{13}{\mega\hertz}) \cite{Ruf2019}.
Additionally, as opposed to NV centres, our V2 centres in \SI{2}{\micro\meter} thick membranes show no signs of spectral diffusion, even at high pump powers.

We can also compare V2 centres in our thinnest \SIrange{0.13}{0.25}{\micro\meter} membrane to Germanium vacancy (GeV) centres in \SI{0.11}{\micro\meter} thin diamond membranes \cite{Guo2021}.
In such membranes, our V2 centres showed linewidths of \SI{\sim150}{\mega\hertz}, which is only \SIrange{6}{11}{} times larger than the lifetime limit, depending on the used optical transition.
For GeV centres, improved spectral stabilities are expected since their structural inversion symmetry isolates them from electric field fluctuations. Indeed, measurements showed an average single-scan linewidths of \SI{95}{\mega\hertz}, which is \SI{\sim4}{} times larger compared to the lifetime limited value of \SI{26}{\mega\hertz} \cite{Bhaskar2017}.
Similarly, tin-vacancy (SnV) inversion symmetry centres in diamond show very good spectral stability even below \SI{60}{\nano\meter} surface proximity \cite{Martinez2022}.
For an individual SnV centre in a photonic waveguide, spectral stability measurements revealed single-scan linewidths of \SI{35}{\mega\hertz}.
This exceeds the lifetime limit only by a factor of $1.63$, even in the presence of off-resonant charge state repumping pulses.
In summary, while the spectral properties of V2 centres in SiC are slightly more sensitive compared to inversion symmetry defects in diamond, they are still surprisingly robust considering the lack of inversion symmetry.
At this point, we mention that this wandering can be counteracted by our previously developed resonance feedback system \cite{Babin2022}, integration of SiC colour centres into p-i-n diodes  for stabilising their charge environments \cite{Anderson2019}, and that we have been able to find several V$_\mathrm{Si}^{-}$ centres with only weak spectral wandering in our membrane D2 (\SIrange{0.63}{0.72}{\micro\meter}, see Supplementary Note 11).

V$_\mathrm{Si}^{-}$ centres in such sub-\SI{}{\micro\meter}-thin membranes are ideally suited for integration into tunable microcavities \cite{Koerber2023}, which can significantly increase photon detection rates via emission funnelling, as well as Purcell enhancement.
With the latter comes also an increase of the natural linewidths of the A$_1$ and A$_2$ optical transitions, which could mitigate spectral wandering to some extent.
Higher photon detection rates will be beneficial for improving detection rates in photonic interference experiments \cite{Nagy2019}, as well as improving the efficiency of the spin state readout for controlling more nuclear spin qubits \cite{Babin2022}.
Since V$_\mathrm{Si}^{-}$ centres can be operated without degradation of spin-optical coherences up to temperatures of $T\sim \SI{20}{\kelvin}$ \cite{Udvarhelyi2020}, an integration into bulky microcavities imposes relatively relaxed constraints on cryogenic cooling power and management.
This qualifies V$_\mathrm{Si}^{-}$ centres in SiC membranes as one of the prime candidates for developing quantum network architectures based on cavity-integrated solid state colour centres. \cite{Heupel2020}


\section*{METHODS}\label{sec:methods}
\subsection*{Sample growth}
The sample of this study is fabricated from an a-plane, n-type 4H-SiC wafer (\textit{Wolfspeed}).
After an external polishing (\textit{NovaSiC}) a \SI{\sim 10}{\micro\meter}-thick n-type epilayer with a free carrier concentration of \SI{\sim 7e13}{\per\centi\meter\cubed} is grown on the polished side along the a-axis by chemical vapour deposition.
After epitaxial growth, the wafer is cut into $5\times \SI{5}{\milli\meter\squared}$ pieces, and the membranes are fabricated within the epilayer of the sample.

\subsection*{Creation of V$_\mathrm{Si}^{-}$ centres}
 For the creation of V$_\mathrm{Si}^{-}$ centres we irradiate the sample with electrons at an energy of \SI{5}{\mega\electronvolt} and a dose of \SI{2}{\kilo\gray}.
 Subsequently, the sample is annealed for \SI{30}{\minute} at \SI{600}{\celsius} (temperature rise of \SI{5}{\degreeCelsius \per \minute} in a $\sim \SI{850}{\milli\bar}$ Ar atmosphere.

\subsection*{Lapping and CMP}
We use a commercial CMP-tool (\textit{PM6, Logitech}) for lapping and polishing of the wafer side.
Lapping is performed with an iron plate, diamond slurry with a mean particle size of \SI{6}{\micro\meter} and a plate rotation speed of \SI{40}{\rpm} at a pressure of \SI{30}{\kilo\pascal}.
Typical removal rates are \SIrange{7}{8}{\micro\meter\per\minute} for this process.
The next step is a rough polishing at the same parameters as the lapping but with a polyurethane pad (\textit{0CON-399, Logitech}) instead of the iron plate with a removal rate of \SI{\sim 1}{\micro\meter\per\minute}.
The final polishing steps are performed with a diamond slurry with a mean particle size of \SI{0.5}{\micro\meter}, for one hour at a pressure of \SI{75}{\kilo\pascal} and for another hour with a softer pad (\textit{0CON-387-1, Logitech}) at a pressure of \SI{150}{\kilo\pascal} with no measurable removal at the \SI{}{\micro\meter}-scale. Our polishing process leads to a small thickness inhomogeneity of the resulting samples due to an edge rounding that becomes stronger with increasing polishing time (see Supplementary Note 12 for details). This is the main reason for the different thicknesses of the individual membranes.

\subsection*{Reactive-ion etching}
The dry etching processes we use are performed in a commercial ICP-RIE tool (\textit{PlasmaPro80, Oxford Instruments}).
It starts with a \SI{2.5}{\minute} O$_2$-plasma soft-ICP cleaning step with a gas flow of \SI{30}{sccm} at a pressure of \SI{10}{\milli\torr}, an ICP power of \SI{180}{\watt}, and a temperature of \SI{24}{\celsius}.
This step is followed by multiple \SI{2}{\minute} SF$_6$-plasma etching steps with a flow of \SI{40}{sccm} at \SI{8}{\milli\torr}, an HF power of \SI{100}{\watt} (no ICP), and a temperature of \SI{20}{\degreeCelsius} with \SI{2}{\minute} waiting steps in between them to prevent a significant heating of the sample.
Typical etch rates for the process are \SIrange{56}{60}{\nano\meter\per\minute}.
The post-fabrication etching step consists of a \SI{2}{\minute} SF$_6$ soft-ICP etching step with a gas flow of \SI{30}{sccm} at a pressure of \SI{10}{\milli\torr}, an ICP power of \SI{250}{\watt}, and a temperature of \SI{24}{\celsius}.

\subsection*{AFM measurements and evaluation}
For surface characterisation we use a commercial AFM (\textit{Veeco Dimension
Icon, Veeco Instruments}) with silicon cantilevers (\textit{Olympus OMCL-AC240TS-R3}) and measure areas of $5 \times \SI{5}{\micro\meter\squared}$. To deduce the RMS-surface roughness, we apply step-line corrections and a polynomial fit on the acquired raw data with the software \textit{Gwyddion}.

\subsection*{Spectra measurements}
We perform the spectra measurements on our sample in a \textit{Montana Instruments} closed-cycle helium cooled cryostat at \SI{\sim 8}{\kelvin} with a home-built confocal microscope using a NA $0.9$ objective  with $\times 100$ magnification and a working distance of \SI{1}{\milli\meter} (\textit{Zeiss EC Epiplan-Neofluar}). To control our hardware, we use the software package \textit{Qudi} \cite{qudi}.
For acquisition of the spectra, we excite the selected spots off-resonantly with a \textit{Toptica iBeam-Smart-CD} \SI{728}{\nano\meter} laser diode and record the spectra using an \textit{Ocean Optics NIRQuest} spectrometer with a resolution of \SIrange{0.3}{0.4}{\nano\meter}. The wavelength of the spectrometer is fine-calibrated by a tunable diode laser  \textit{Toptica DL pro} and a wavemeter \textit{HighFinesse Angstrom WS7-60}. The peaks from the spectra are extracted using the \textit{SciPy} method \textit{find\_peaks}.

\subsection*{PLE measurements}
We perform the PLE measurements on the same low-temperature confocal microscope setup and use the same off-resonant repump laser as in the spectra measurements.
To excite an emitter resonantly, we utilise a tunable \textit{Toptica DL pro} laser whose light passes an acousto-optic modulator by \textit{Gooch\&Housego} operating at \SI{270}{\mega\hertz} that acts as a switch.
The excitation power of the laser is measured before the objective.
We filter out photons emitted in the ZPL using two tunable long pass filters \textit{Semrock TLP01-995} and detect photons in the PSB before the post-fabrication with an APD \textit{SPCM-900-24} by \textit{Excelitas Technologies Corp} and after the post-fabrication with an SNSPD by \textit{Photon Spot}.
The laser frequency is monitored using a wavemeter \textit{HighFinesse Angstrom WS7-60} and tuned by applying a voltage via a \textit{National Instruments} data acquisition card (NIDAQ).
A PLE line consists of the collected counts with respect to this applied voltage.
The typical tuning range amounts to \SI{\sim2}{\giga\hertz} around a wavelength of \SI{917}{\nano\meter}.
To avoid decaying into a dark state via inter-system-crossing after excitation, we drive the ground state spin transition continuously using a microwave signal at input powers of \SIrange{12}{15}{dBm}.
Only in case the system ionises in a single line scan, we turn on a \SI{728}{\nano\meter} laser (software-wise implemented as pulses at intervals of \SI{1}{\milli\second} for \SI{10}{\micro\second} each), to repump the system back to its negatively charged state, in the time the tunable laser returns to its starting frequency.
One can determine a linewidth from a PLE scan consisting of multiple PLE lines, by summing all lines, performing a Lorentzian fit, and transforming the FWHM from voltage to frequency based on the known separation of the A$_1$ and A$_2$ transition peaks (\SI{1}{\giga\hertz}, \cite{Nagy2021}).
However, this method only works for scans with little to no observable spectral wandering, since more than two peaks can appear in case of a stronger spectral wandering, see the middle panel of Figure \ref{Fig3}(c).
For this purpose, we have written a program that starts by trying to fit each single line scan to the sum of two Lorentzian profiles using the \textit{SciPy} method \textit{curve\_fit} with restrains on the fit parameters, such as no negative amplitudes, FWHMs smaller than the scanned interval, manually defined limits to the peak positions, and only a small variation in the distance between the peak positions.
Next, it shifts the lines where this fit was successful according to the fit parameters such that the positions of all A$_1$ (A$_2$) peaks align, sums the shifted lines and performs another Lorentzian fit.
Only then is the data transformed to frequencies based on the peak splitting and the A$_1$ (A$_2$) linewidth extracted.
Since a standard deviation of \SI{75}{\mega\hertz} for the ZFS of the excited state was reported for V1 centres \cite{Nagy2021}, and we expect a similar behaviour for V2 centres, we attribute an error of \SI{7.5}{\percent} to our extracted linewidths.
The spectral wandering in the histogram in Figure \ref{Fig4} stems from the peak position difference in single line fits for two consecutive lines divided by the time in between those lines.
This data is converted from voltages to frequency based on the transformation value obtained as described above.
We have calculated the mean errors for each evaluated region and chose the bin width to be the biggest one of these, namely \SI{2.13}{\mega\hertz\per\second}.
An insignificant amount of data points (8 out of 3471) was disregarded in the histogram, since the obtained standard deviation of the single line fits used for those points is unphysically high (\SI{>200}{\mega\hertz\per\second}).

\section*{DATA AVAILABILITY}
The data supporting the presented findings are available at the following repository: \url{https://doi.org/10.18419/darus-3982}.

\section*{CODE AVAILABILITY}
The measurement and evaluation codes used for this study are available from the corresponding author upon reasonable request.

\section*{AUTHOR CONTRIBUTIONS}
J.H., J.K., and E.H. contributed equally to this work.
The project was conceived by J.H., J.K., P.F., C.B., and F.K. and supervised by F.K. and J.W.
The high-quality epilayer was grown by J.U.H. and M.G. and the sample was electron irradiated by W.K.
J.H., J.K., R.S., and P.F. have been involved in the fabrication and related measurements of all shown samples.
The optical setups were constructed by J.H., J.K., E.H., P.K., P.F., and F.K.
The optical measurements were conducted by J.H., E.H., and P.K. and analysed by J.H., J.K., E.H., and F.K.
The manuscript was written by J.H., J.K., and F.K.
All authors contributed to the manuscript.

\section*{COMPETING INTERESTS}
The authors declare no competing interests.

\section*{ACKNOWLEDGEMENTS}
We acknowledge fruitful discussions with Dr. Daniil Lukin, Dr. Melissa Guidry, Dr. Julia Heupel, Dr. Roman Kolesov, Vladislav Bushmakin, and Marcel Krumrein, experimental help from Arnold Weible and Ole Sumpf as well as assistance in proof-reading of the manuscript from Raphael Wörnle.\\
J.H., F.K., and J.W. acknowledge support from the European Commission for the Quantum Technology Flagship project QIA (Grant agreements No. 101080128, and 101102140).
P.K., F.K., J.U.H, and J.W. acknowledge support from the European Commission through the QuantERA project InQuRe (Grant agreements No. 731473, and 101017733).
P.K., F.K. and J.W. acknowledge the German ministry of education and research for the project InQuRe (BMBF, Grant agreement No. 16KIS1639K).
C.B., F.K. and J.W. further acknowledge the German ministry of education and research for the project QR.X (BMBF, Grant agreements No. 16KISQ001K and 16KISQ013), while J.W. also acknowledges support for the project Spinning (BMBF, Grant agreement No. 13N16219).
F.K. and J.W. additionally acknowledge the Baden-Württemberg Stiftung for the project SPOC (Grant agreement No. QT-6).
F.K. acknowledges funding by the Luxembourg National Research Fund (FNR, project: 17792569) in addition.
J.U.H further acknowledges support from the Swedish Research Council under VR Grant No. 2020-05444 and Knut and Alice Wallenberg Foundation (Grant No. KAW 2018.0071).

\EndMatter

@article{Lukin2020,
    title = {{4H-silicon-carbide-on-insulator for integrated quantum and nonlinear photonics}},
    year = {2020},
    journal = {Nat. Photonics},
    author = {Lukin, Daniil M. and Dory, Constantin and Guidry, Melissa A. and Yang, Ki Youl and Mishra, Sattwik Deb and Trivedi, Rahul and Radulaski, Marina and Sun, Shuo and Vercruysse, Dries and Ahn, Geun Ho and Vu{\v{c}}kovi{\'{c}}, Jelena},
    number = {5},
    month = {12},
    pages = {330--334},
    volume = {14},
    publisher = {Nature Publishing Group},
    url = {https://www.nature.com/articles/s41566-019-0556-6},
    doi = {10.1038/s41566-019-0556-6},
    issn = {17494893}
}

@article{Fuchs2021,
    title = {{A cavity-based optical antenna for color centers in diamond}},
    year = {2021},
    journal = {APL Photonics},
    author = {Fuchs, Philipp and Jung, Thomas and Kieschnick, Michael and Meijer, Jan and Becher, Christoph},
    number = {},
    month = {8},
    pages = {086102},
    volume = {6},
    publisher = {American Institute of Physics Inc.},
    doi = {10.1063/5.0057161},
    issn = {23780967},
    arxivId = {2105.10249}
}

@article{Bradley2019,
    title = {{A Ten-Qubit Solid-State Spin Register with Quantum Memory up to One Minute}},
    year = {2019},
    journal = {Phys. Rev. X},
    author = {Bradley, C. E. and Randall, J. and Abobeih, M. H. and Berrevoets, R. C. and Degen, M. J. and Bakker, M. A. and Markham, M. and Twitchen, D. J. and Taminiau, T. H.},
    number = {3},
    month = {9},
    pages = {031045},
    volume = {9},
    publisher = {American Physical Society},
    url = {https://journals.aps.org/prx/abstract/10.1103/PhysRevX.9.031045},
    doi = {10.1103/PhysRevX.9.031045},
    issn = {21603308},
    arxivId = {1905.02094}
}

@article{DurandSi2021,
    title = {{Broad Diversity of Near-Infrared Single-Photon Emitters in Silicon}},
    year = {2021},
    journal = {Phys. Rev. Lett.},
    author = {Durand, A. and Baron, Y. and Redjem, W. and Herzig, T. and Benali, A. and Pezzagna, S. and Meijer, J. and Kuznetsov, A. Yu and G{\'{e}}rard, J. M. and Robert-Philip, I. and Abbarchi, M. and Jacques, V. and Cassabois, G. and Dr{\'{e}}au, A.},
    number = {},
    month = {2},
    pages = {083602},
    volume = {126},
    publisher = {American Physical Society},
    doi = {10.1103/PhysRevLett.126.083602},
    issn = {10797114},
    pmid = {33709758},
    arxivId = {2010.11068}
}

@article{Wang2023,
    title = {{Chemical–Mechanical Polishing of 4H Silicon Carbide Wafers}},
    year = {2023},
    journal = {Adv. Mater. Interfaces},
    author = {Wang, Wantang and Lu, Xuesong and Wu, Xinke and Zhang, Yiqiang and Wang, Rong and Yang, Deren and Pi, Xiaodong},
    number = {13},
    month = {5},
    pages = {2202369},
    volume = {10},
    publisher = {John Wiley {\&} Sons, Ltd},
    url = {https://onlinelibrary.wiley.com/doi/full/10.1002/admi.202202369 https://onlinelibrary.wiley.com/doi/abs/10.1002/admi.202202369 https://onlinelibrary.wiley.com/doi/10.1002/admi.202202369},
    doi = {10.1002/ADMI.202202369},
    issn = {2196-7350}
}

@article{Salz2020,
    title = {{Cryogenic platform for coupling color centers in diamond membranes to a fiber-based microcavity}},
    year = {2020},
    journal = {Appl. Phys. B},
    author = {Salz, M. and Herrmann, Y. and Nadarajah, A. and Stahl, A. and Hettrich, M. and Stacey, A. and Prawer, S. and Hunger, D. and Schmidt-Kaler, F.},
    number = {8},
    month = {8},
    pages = {131},
    volume = {126},
    publisher = {Springer},
    url = {https://link.springer.com/article/10.1007/s00340-020-07478-5},
    doi = {10.1007/s00340-020-07478-5},
    issn = {09462171},
    arxivId = {2002.08304}
}

@article{Riedel2017,
    title = {{Deterministic enhancement of coherent photon generation from a nitrogen-vacancy center in ultrapure diamond}},
    year = {2017},
    journal = {Phys. Rev. X},
    author = {Riedel, Daniel and S{\"{o}}llner, Immo and Shields, Brendan J. and Starosielec, Sebastian and Appel, Patrick and Neu, Elke and Maletinsky, Patrick and Warburton, Richard J.},
    number = {3},
    month = {9},
    pages = {031040},
    volume = {7},
    publisher = {American Physical Society},
    url = {https://journals.aps.org/prx/abstract/10.1103/PhysRevX.7.031040},
    doi = {10.1103/PhysRevX.7.031040},
    issn = {21603308},
    arxivId = {1703.00815}
}

@article{Haeussler2019,
    title = {{Diamond photonics platform based on silicon vacancy centers in a single-crystal diamond membrane and a fiber cavity}},
    year = {2019},
    journal = {Phys. Rev. B},
    author = {H{\"{a}}u{\ss}ler, Stefan and Benedikter, Julia and Bray, Kerem and Regan, Blake and Dietrich, Andreas and Twamley, Jason and Aharonovich, Igor and Hunger, David and Kubanek, Alexander},
    number = {16},
    month = {4},
    pages = {165310},
    volume = {99},
    publisher = {American Physical Society},
    url = {https://journals.aps.org/prb/abstract/10.1103/PhysRevB.99.165310},
    doi = {10.1103/PhysRevB.99.165310},
    issn = {24699969},
    arxivId = {1812.02426}
}

@article{Anderson2019,
    title = {{Electrical and optical control of single spins integrated in scalable semiconductor devices}},
    year = {2019},
    journal = {Science},
    author = {Anderson, Christopher P. and Bourassa, Alexandre and Miao, Kevin C. and Wolfowicz, Gary and Mintun, Peter J. and Crook, Alexander L. and Abe, Hiroshi and Ul Hassan, Jawad and Son, Nguyen T. and Ohshima, Takeshi and Awschalom, David D.},
    number = {6470},
    month = {12},
    pages = {1225--1230},
    volume = {366},
    publisher = {American Association for the Advancement of Science},
    url = {https://www.science.org/doi/10.1126/science.aax9406},
    doi = {10.1126/SCIENCE.AAX9406/SUPPL{\_}FILE/AAX9406-ANDERSON-SM.PDF},
    issn = {10959203},
    pmid = {31806809},
    arxivId = {1906.08328}
}

@article{Wagner2000,
    title = {{Electronic structure of the neutral silicon vacancy in 4H and 6H SiC}},
    year = {2000},
    journal = {Phys. Rev. B},
    author = {Wagner, Mt and Magnusson, B. and Chen, W. M. and Janz{\'{e}}n, E. and S{\"{o}}rman, E. and Hallin, C. and Lindstrom, J. L.},
    number = {},
    month = {12},
    pages = {016555},
    volume = {62},
    publisher = {American Physical Society},
    url = {https://journals.aps.org/prb/abstract/10.1103/PhysRevB.62.16555},
    doi = {10.1103/PhysRevB.62.16555},
    issn = {01631829}
}

@article{BishopGaN2022,
    title = {{Enhanced light collection from a gallium nitride color center using a near index-matched solid immersion lens}},
    year = {2022},
    journal = {Appl. Phys. Lett.},
    author = {Bishop, S. G. and Hadden, J. P. and Hekmati, R. and Cannon, J. K. and Langbein, W. W. and Bennett, A. J.},
    number = {},
    month = {3},
    pages = {114001},
    volume = {120},
    publisher = {American Institute of Physics Inc.},
    doi = {10.1063/5.0085257},
    issn = {00036951},
    arxivId = {2202.06754}
}

@article{NiaourisZnO2022,
    title = {{Ensemble spin relaxation of shallow donor qubits in ZnO}},
    year = {2022},
    journal = {Phys. Rev. B},
    author = {Niaouris, Vasileios and Durnev, Mikhail V. and Linpeng, Xiayu and Viitaniemi, Maria L.K. and Zimmermann, Christian and Vishnuradhan, Aswin and Kozuka, Yusuke and Kawasaki, Masashi and Fu, Kai Mei C.},
    number = {},
    month = {5},
    pages = {195202},
    volume = {105},
    publisher = {American Physical Society},
    doi = {10.1103/PhysRevB.105.195202},
    issn = {24699969},
    arxivId = {2111.11564}
}

@article{Heupel2020,
    title = {{Fabrication and characterization of single-crystal diamond membranes for quantum photonics with tunable microcavities}},
    year = {2020},
    journal = {Micromachines},
    author = {Heupel, Julia and Pallmann, Maximilian and K{\"{o}}rber, Jonathan and Merz, Rolf and Kopnarski, Michael and St{\"{o}}hr, Rainer and Reithmaier, Johann Peter and Hunger, David and Popov, Cyril},
    number = {12},
    month = {12},
    pages = {1--18},
    volume = {11},
    publisher = {Multidisciplinary Digital Publishing Institute},
    url = {https://www.mdpi.com/2072-666X/11/12/1080/htm https://www.mdpi.com/2072-666X/11/12/1080},
    doi = {10.3390/mi11121080},
    issn = {2072666X}
}

@article{Babin2022,
    title = {{Fabrication and nanophotonic waveguide integration of silicon carbide colour centres with preserved spin-optical coherence}},
    year = {2022},
    journal = {Nat. Mater.},
    author = {Babin, Charles and St{\"{o}}hr, Rainer and Morioka, Naoya and Linkewitz, Tobias and Steidl, Timo and W{\"{o}}rnle, Raphael and Liu, Di and Hesselmeier, Erik and Vorobyov, Vadim and Denisenko, Andrej and Hentschel, Mario and Gobert, Christian and Berwian, Patrick and Astakhov, Georgy V. and Knolle, Wolfgang and Majety, Sridhar and Saha, Pranta and Radulaski, Marina and Son, Nguyen Tien and Ul-Hassan, Jawad and Kaiser, Florian and Wrachtrup, Jörg},
    number = {1},
    month = {11},
    pages = {67--73},
    volume = {21},
    publisher = {Nature Publishing Group},
    url = {https://www.nature.com/articles/s41563-021-01148-3},
    doi = {10.1038/s41563-021-01148-3},
    issn = {14764660},
    pmid = {34795400}
}

@article{Anderson5s2022,
    title = {{Five-second coherence of a single spin with single-shot readout in silicon carbide}},
    year = {2022},
    journal = {Sci. Adv.},
    author = {Anderson, Christopher P and Glen, Elena O and Zeledon, Cyrus and Bourassa, Alexandre and Jin, Yu and Zhu, Yizhi and Vorwerk, Christian and Crook, Alexander L and Abe, Hiroshi and Ul-Hassan, Jawad and Ohshima, Takeshi and Son, Nguyen T and Galli, Giulia and Awschalom, David D},
    pages = {5912},
    volume = {8},
    url = {https://www.science.org}
}

@article{Nagy2019,
    title = {{High-fidelity spin and optical control of single silicon-vacancy centres in silicon carbide}},
    year = {2019},
    journal = {Nat. Commun.},
    author = {Nagy, Roland and Niethammer, Matthias and Widmann, Matthias and Chen, Yu-Chen and Udvarhelyi, Péter and Bonato, Cristian and Hassan, Jawad Ul and Karhu, Robin and Ivanov, Ivan G and Son, Nguyen Tien and Maze, Jeronimo R and Ohshima, Takeshi and Soykal, Öney O and Gali, Ádám and Lee, Sang-Yun and Kaiser, Florian and Wrachtrup, Jörg},
    number = {1},
    pages = {1954},
    volume = {10},
    url = {https://doi.org/10.1038/s41467-019-09873-9},
    doi = {10.1038/s41467-019-09873-9},
    issn = {2041-1723}
}

@article{Ivady2017,
    title = {{Identification of Si-vacancy related room-temperature qubits in 4H silicon carbide}},
    year = {2017},
    journal = {Phys. Rev. B},
    author = {Iv{\'{a}}dy, Viktor and Davidsson, Joel and Son, Nguyen Tien and Ohshima, Takeshi and Abrikosov, Igor A and Gali, Adam},
    number = {16},
    pages = {161114},
    volume = {96},
    doi = {10.1103/PhysRevB.96.161114}
}

@article{Geng2022,
    title = {{Identification of subsurface damage of 4H-SiC wafers by combining photo-chemical etching and molten-alkali etching}},
    year = {2022},
    journal = {J. Semicond.},
    author = {Geng, Wenhao and Yang, Guang and Zhang, Xuqing and Zhang, Xi and Wang, Yazhe and Song, Lihui and Chen, Penglei and Zhang, Yiqiang and Pi, Xiaodong and Yang, Deren},
    number = {10},
    month = {10},
    pages = {102801},
    volume = {43},
    publisher = {IOP Publishing},
    url = {https://iopscience.iop.org/article/10.1088/1674-4926/43/10/102801 https://iopscience.iop.org/article/10.1088/1674-4926/43/10/102801/meta},
    doi = {10.1088/1674-4926/43/10/102801},
    issn = {1674-4926}
}

@article{Nagy2021,
    title = {{Narrow inhomogeneous distribution of spin-active emitters in silicon carbide}},
    year = {2021},
    journal = {Appl. Phys. Lett.},
    author = {Nagy, Roland and Dasari, Durga Bhaktavatsala Rao and Babin, Charles and Liu, Di and Vorobyov, Vadim and Niethammer, Matthias and Widmann, Matthias and Linkewitz, Tobias and Gediz, Izel and St{\"{o}}hr, Rainer and Weber, Heiko B. and Ohshima, Takeshi and Ghezellou, Misagh and Son, Nguyen Tien and Ul-Hassan, Jawad and Kaiser, Florian and Wrachtrup, Jörg},
    number = {14},
    month = {4},
    pages = {144003},
    volume = {118},
    publisher = { AIP Publishing LLC AIP Publishing },
    url = {https://aip.scitation.org/doi/abs/10.1063/5.0046563},
    doi = {10.1063/5.0046563},
    issn = {0003-6951},
    arxivId = {2103.06101}
}

@article{HigginbottomSi2022,
    title = {{Optical observation of single spins in silicon}},
    year = {2022},
    journal = {Nature},
    author = {Higginbottom, Daniel B. and Kurkjian, Alexander T.K. and Chartrand, Camille and Kazemi, Moein and Brunelle, Nicholas A. and MacQuarrie, Evan R. and Klein, James R. and Lee-Hone, Nicholas R. and Stacho, Jakub and Ruether, Myles and Bowness, Camille and Bergeron, Laurent and DeAbreu, Adam and Harrigan, Stephen R. and Kanaganayagam, Joshua and Marsden, Danica W. and Richards, Timothy S. and Stott, Leea A. and Roorda, Sjoerd and Morse, Kevin J. and Thewalt, Michael L.W. and Simmons, Stephanie},
    number = {},
    month = {7},
    pages = {266},
    volume = {607},
    publisher = {Nature Research},
    doi = {10.1038/s41586-022-04821-y},
    issn = {14764687},
    pmid = {35831600},
    arxivId = {2103.07580}
}

@article{Ruf2019,
    title = {{Optically Coherent Nitrogen-Vacancy Centers in Micrometer-Thin Etched Diamond Membranes}},
    year = {2019},
    journal = {Nano Lett.},
    author = {Ruf, Maximilian and Ijspeert, Mark and Van Dam, Suzanne and De Jong, Nick and Van Den Berg, Hans and Evers, Guus and Hanson, Ronald},
    number = {6},
    month = {6},
    pages = {3987--3992},
    volume = {19},
    publisher = {American Chemical Society},
    url = {https://pubs.acs.org/doi/full/10.1021/acs.nanolett.9b01316},
    doi = {10.1021/ACS.NANOLETT.9B01316/SUPPL{\_}FILE/NL9B01316{\_}SI{\_}001.PDF},
    issn = {15306992},
    pmid = {31136192},
    arxivId = {1904.00883}
}

@article{Orphal2023,
    title = {{Optically Coherent Nitrogen-Vacancy Defect Centers in Diamond Nanostructures}},
    year = {2023},
    journal = {Phys. Rev. X},
    author = {Orphal-Kobin, Laura and Unterguggenberger, Kilian and Pregnolato, Tommaso and Kemf, Natalia and Matalla, Mathias and Unger, Ralph Stephan and Ostermay, Ina and Pieplow, Gregor and Schr{\"{o}}der, Tim},
    number = {1},
    month = {1},
    pages = {011042},
    volume = {13},
    publisher = {American Physical Society},
    url = {https://journals.aps.org/prx/abstract/10.1103/PhysRevX.13.011042},
    doi = {10.1103/PHYSREVX.13.011042/FIGURES/14/MEDIUM},
    issn = {21603308},
    arxivId = {2203.05605}
}

@article{Martinez2022,
    title = {{Photonic Indistinguishability of the Tin-Vacancy Center in Nanostructured Diamond}},
    year = {2022},
    journal = {Phys. Rev. Lett.},
    author = {Arjona Mart{\'{i}}nez, Jesús and Parker, Ryan A. and Chen, Kevin C. and Purser, Carola M. and Li, Linsen and Michaels, Cathryn P. and Stramma, Alexander M. and Debroux, Romain and Harris, Isaac B. and Hayhurst Appel, Martin and Nichols, Eleanor C. and Trusheim, Matthew E. and Gangloff, Dorian A. and Englund, Dirk and Atat{\"{u}}re, Mete},
    number = {17},
    month = {10},
    pages = {173603},
    volume = {129},
    publisher = {American Physical Society},
    url = {https://journals.aps.org/prl/abstract/10.1103/PhysRevLett.129.173603},
    doi = {10.1103/PhysRevLett.129.173603},
    issn = {10797114},
    pmid = {36332262},
    arxivId = {2206.15239}
}

@article{CannonAlN2023,
    title = {{Polarization study of single color centers in aluminum nitride}},
    year = {2023},
    journal = {Appl. Phys. Lett.},
    author = {Cannon, J. K. and Bishop, S. G. and Hadden, J. P. and Ya{\u{g}}cı, H. B. and Bennett, A. J.},
    number = {},
    pages = {172104},
    volume = {122},
    publisher = {American Institute of Physics Inc.},
    doi = {10.1063/5.0145542},
    issn = {00036951}
}

@article{Bhaskar2017,
    title = {{Quantum Nonlinear Optics with a Germanium-Vacancy Color Center in a Nanoscale Diamond Waveguide}},
    year = {2017},
    journal = {Phys. Rev. Lett.},
    author = {Bhaskar, M. K. and Sukachev, D. D. and Sipahigil, A. and Evans, R. E. and Burek, M. J. and Nguyen, C. T. and Rogers, L. J. and Siyushev, P. and Metsch, M. H. and Park, H. and Jelezko, F. and Lon{\v{c}}ar, M. and Lukin, M. D.},
    number = {22},
    month = {5},
    pages = {223603},
    volume = {118},
    publisher = {American Physical Society},
    url = {https://journals.aps.org/prl/abstract/10.1103/PhysRevLett.118.223603},
    doi = {10.1103/PHYSREVLETT.118.223603/FIGURES/5/MEDIUM},
    issn = {10797114},
    pmid = {28621982},
    arxivId = {1612.03036}
}

@article{Awschalom2018,
    title = {{Quantum technologies with optically interfaced solid-state spins}},
    year = {2018},
    journal = {Nat. Photonics},
    author = {Awschalom, David D. and Hanson, Ronald and Wrachtrup, Jörg and Zhou, Brian B.},
    number = {9},
    month = {9},
    pages = {516--527},
    volume = {12},
    publisher = {Nature Publishing Group},
    doi = {10.1038/s41566-018-0232-2},
    issn = {17494893}
}

@article{HermansTele2022,
    title = {{Qubit teleportation between non-neighbouring nodes in a quantum network}},
    year = {2022},
    journal = {Nature},
    author = {Hermans, S. L.N. and Pompili, M. and Beukers, H. K.C. and Baier, S. and Borregaard, J. and Hanson, R.},
    number = {},
    month = {5},
    pages = {663},
    volume = {605},
    publisher = {Nature Research},
    doi = {10.1038/s41586-022-04697-y},
    issn = {14764687},
    pmid = {35614248},
    arxivId = {2110.11373}
}

@article{qudi,
    title = {{Qudi: A modular python suite for experiment control and data processing}},
    year = {2017},
    journal = {SoftwareX},
    author = {Binder, Jan M. and Stark, Alexander and Tomek, Nikolas and Scheuer, Jochen and Frank, Florian and Jahnke, Kay D. and M{\"{u}}ller, Christoph and Schmitt, Simon and Metsch, Mathias H. and Unden, Thomas and Gehring, Tobias and Huck, Alexander and Andersen, Ulrik L. and Rogers, Lachlan J. and Jelezko, Fedor},
    month = {1},
    pages = {85--90},
    volume = {6},
    publisher = {Elsevier},
    doi = {10.1016/J.SOFTX.2017.02.001},
    issn = {2352-7110},
    arxivId = {1611.09146}
}

@article{Ruf2021,
    title = {{Resonant Excitation and Purcell Enhancement of Coherent Nitrogen-Vacancy Centers Coupled to a Fabry-Perot Microcavity}},
    year = {2021},
    journal = {Phys. Rev. Applied},
    author = {Ruf, M. and Weaver, M. J. and Van Dam, S. B. and Hanson, R.},
    number = {2},
    month = {2},
    pages = {024049},
    volume = {15},
    publisher = {American Physical Society},
    url = {https://journals.aps.org/prapplied/abstract/10.1103/PhysRevApplied.15.024049},
    doi = {10.1103/PhysRevApplied.15.024049},
    issn = {23317019},
    arxivId = {2009.08204}
}

@article{Banks2019,
    title = {{Resonant Optical Spin Initialization and Readout of Single Silicon Vacancies in 4H - Si C}},
    year = {2019},
    journal = {Phys. Rev. Applied},
    author = {Banks, Hunter B. and Soykal, Öney O. and Myers-Ward, Rachael L. and Gaskill, D. Kurt and Reinecke, T. L. and Carter, Samuel G.},
    number = {2},
    month = {2},
    pages = {024013},
    volume = {11},
    publisher = {American Physical Society},
    doi = {10.1103/PhysRevApplied.11.024013},
    issn = {23317019},
    arxivId = {1811.01293}
}

@article{LuoGaN2023,
    title = {{Room temperature optically detected magnetic resonance of single spins in GaN}},
    year = {2024},
    journal = {Nat. Mater.},
    author = {Luo, Jialun and Geng, Yifei and Rana, Farhan and Fuchs, Gregory D.},
    doi = {10.1038/s41563-024-01803-5}
}

@article{Radulaski2017,
    title = {{Scalable Quantum Photonics with Single Color Centers in Silicon Carbide}},
    year = {2017},
    journal = {Nano Lett.},
    author = {Radulaski, Marina and Widmann, Matthias and Niethammer, Matthias and Zhang, Jingyuan Linda and Lee, Sang Yun and Rendler, Torsten and Lagoudakis, Konstantinos G. and Son, Nguyen Tien and Janz{\'{e}}n, Erik and Ohshima, Takeshi and Wrachtrup, Jörg and Vu{\v{c}}kovi{\'{c}}, Jelena},
    number = {3},
    month = {3},
    pages = {1782--1786},
    volume = {17},
    publisher = {American Chemical Society},
    url = {https://pubs.acs.org/doi/full/10.1021/acs.nanolett.6b05102},
    doi = {10.1021/acs.nanolett.6b05102},
    issn = {15306992},
    pmid = {28225630},
    arxivId = {1612.02874}
}

@article{Koerber2023,
    title = {{Scanning Cavity Microscopy of a Single-Crystal Diamond Membrane}},
    year = {2023},
    journal = {Phys. Rev. Applied},
    author = {K{\"{o}}rber, Jonathan and Pallmann, Maximilian and Heupel, Julia and St{\"{o}}hr, Rainer and Vasilenko, Evgenij and H{\"{u}}mmer, Thomas and Kohler, Larissa and Popov, Cyril and Hunger, David},
    number = {6},
    month = {6},
    pages = {064057},
    volume = {19},
    publisher = {American Physical Society},
    url = {https://journals.aps.org/prapplied/abstract/10.1103/PhysRevApplied.19.064057},
    doi = {10.1103/PhysRevApplied.19.064057},
    issn = {2331-7019}
}

@article{Castelletto2020,
    title = {{Silicon carbide color centers for quantum applications}},
    year = {2020},
    journal = {J. Phys. Photonics},
    author = {Castelletto, Stefania and Boretti, Alberto},
    number = {},
    month = {3},
    pages = {022001},
    volume = {2},
    publisher = {IOP Publishing Ltd},
    doi = {10.1088/2515-7647/ab77a2},
    issn = {25157647}
}

@article{Castelletto2021,
    title = {{Silicon Carbide Photonics Bridging Quantum Technology}},
    year = {2021},
    journal = {ACS Photonics},
    author = {Castelletto, Stefania and Peruzzo, Alberto and Bonato, Cristian and Johnson, Brett C. and Radulaski, Marina and Ou, Haiyan and Kaiser, Florian and Wrachtrup, Joerg},
    pages = {1434--1457},
    volume = { 9},
    publisher = {American Chemical Society},
    url = {https://pubs.acs.org/doi/full/10.1021/acsphotonics.1c01775},
    doi = {10.1021/acsphotonics.1c01775},
    issn = {23304022}
}

@article{Lukin2020Stark,
    title = {{Spectrally reconfigurable quantum emitters enabled by optimized fast modulation}},
    year = {2020},
    journal = {Npj Quantum Inf.},
    author = {Lukin, Daniil M. and White, Alexander D. and Trivedi, Rahul and Guidry, Melissa A. and Morioka, Naoya and Babin, Charles and Soykal, Öney O. and Ul-Hassan, Jawad and Son, Nguyen Tien and Ohshima, Takeshi and Vasireddy, Praful K. and Nasr, Mamdouh H. and Sun, Shuo and MacLean, Jean Philippe W. and Dory, Constantin and Nanni, Emilio A. and Wrachtrup, Jörg and Kaiser, Florian and Vu{\v{c}}kovi{\'{c}}, Jelena},
    number = {1},
    month = {9},
    pages = {80},
    volume = {6},
    publisher = {Nature Publishing Group},
    url = {https://www.nature.com/articles/s41534-020-00310-0},
    doi = {10.1038/s41534-020-00310-0},
    issn = {2056-6387},
    arxivId = {2003.12591}
}

@article{Tarasenko2018,
    title = {{Spin and Optical Properties of Silicon Vacancies in Silicon Carbide - A Review}},
    year = {2018},
    journal = {Phys. Status Solidi B},
    author = {Tarasenko, S. A. and Poshakinskiy, A. V. and Simin, D. and Soltamov, V. A. and Mokhov, E. N. and Baranov, P. G. and Dyakonov, V. and Astakhov, G. V.},
    number = {1},
    month = {1},
    pages = {1700258},
    volume = {255},
    publisher = {Wiley-VCH Verlag},
    doi = {10.1002/pssb.201700258},
    issn = {15213951},
    arxivId = {1707.05503}
}

@article{Morioka2020,
    title = {{Spin-controlled generation of indistinguishable and distinguishable photons from silicon vacancy centres in silicon carbide}},
    year = {2020},
    journal = {Nat. Commun.},
    author = {Morioka, Naoya and Babin, Charles and Nagy, Roland and Gediz, Izel and Hesselmeier, Erik and Liu, Di and Joliffe, Matthew and Niethammer, Matthias and Dasari, Durga and Vorobyov, Vadim and Kolesov, Roman and St{\"{o}}hr, Rainer and Ul-Hassan, Jawad and Son, Nguyen Tien and Ohshima, Takeshi and Udvarhelyi, Péter and Thiering, Gergő and Gali, Adam and Wrachtrup, Jörg and Kaiser, Florian},
    number = {},
    month = {5},
    pages = {2516},
    volume = {11},
    publisher = {Nature Publishing Group},
    url = {https://www.nature.com/articles/s41467-020-16330-5},
    doi = {10.1038/s41467-020-16330-5},
    issn = {2041-1723},
    pmid = {32433556},
    arxivId = {2001.02455}
}

@article{Ruehl2020,
    title = {{Stark Tuning of the Silicon Vacancy in Silicon Carbide}},
    year = {2020},
    journal = {Nano Lett.},
    author = {R{\"{u}}hl, Maximilian and Bergmann, Lena and Krieger, Michael and Weber, Heiko B.},
    number = {1},
    month = {1},
    pages = {658--663},
    volume = {20},
    publisher = {American Chemical Society},
    url = {https://pubs.acs.org/doi/pdf/10.1021/acs.nanolett.9b04419},
    doi = {10.1021/ACS.NANOLETT.9B04419/ASSET/IMAGES/MEDIUM/NL9B04419{\_}0003.GIF},
    issn = {15306992},
    pmid = {31809057}
}

@article{DiPaper,
    title = {{The silicon vacancy centers in SiC: determination of intrinsic spin dynamics for integrated quantum photonics. Preprint at https://arxiv.org/abs/2307.13648v1}},
    year = {2023},
    author = {Liu, Di and Kaiser, Florian and Bushmakin, Vladislav and Hesselmeier, Erik and Steidl, Timo and Ohshima, Takeshi and Son, Nguyen Tien and Ul-Hassan, Jawad and Soykal, Öney O. and Wrachtrup, Jörg},
    month = {7},
    url = {https://arxiv.org/abs/2307.13648v1},
    arxivId = {2307.13648}
}

@article{Guo2021,
    title = {{Tunable and Transferable Diamond Membranes for Integrated Quantum Technologies}},
    year = {2021},
    journal = {Nano Lett.},
    author = {Guo, Xinghan and Delegan, Nazar and Karsch, Jonathan C. and Li, Zixi and Liu, Tianle and Shreiner, Robert and Butcher, Amy and Awschalom, David D. and Heremans, F. Joseph and High, Alexander A.},
    number = {24},
    month = {12},
    pages = {10392--10399},
    volume = {21},
    publisher = {American Chemical Society},
    url = {https://pubs.acs.org/doi/full/10.1021/acs.nanolett.1c03703},
    doi = {doi/10.1021/acs.nanolett.1c03703},
    issn = {15306992},
    pmid = {34894697},
    arxivId = {2109.11507}
}

@article{Lukin2023,
    title = {{Two-Emitter Multimode Cavity Quantum Electrodynamics in Thin-Film Silicon Carbide Photonics}},
    year = {2023},
    journal = {Phys. Rev. X},
    author = {Lukin, Daniil M. and Guidry, Melissa A. and Yang, Joshua and Ghezellou, Misagh and Deb Mishra, Sattwik and Abe, Hiroshi and Ohshima, Takeshi and Ul-Hassan, Jawad and Vu{\v{c}}kovi{\'{c}}, Jelena},
    number = {1},
    month = {1},
    pages = {011005},
    volume = {13},
    publisher = {American Physical Society},
    url = {https://journals.aps.org/prx/abstract/10.1103/PhysRevX.13.011005},
    doi = {10.1103/PhysRevX.13.011005},
    issn = {21603308}
}

@article{Udvarhelyi2020,
    title = {{Vibronic States and Their Effect on the Temperature and Strain Dependence of Silicon-Vacancy Qubits in 4H-SiC}},
    year = {2020},
    journal = {Phys. Rev. Applied},
    author = {Udvarhelyi, Péter and Morioka, Naoya and Babin, Charles and Kaiser, Florian and Lukin, Daniil and Ohshima, Takeshi and Ul-Hassan, Jawad and Tien Son, Nguyen and Vu{\v{c}}kovi, Jelena and Wrachtrup, Jörg and Gali, Adam},
    pages = {054017},
    volume = {10},
    doi = {10.1103/PhysRevApplied.13.054017}
}

@article{HollenbachSi2022,
    title = {{Wafer-scale nanofabrication of telecom single-photon emitters in silicon}},
    year = {2022},
    journal = {Nat. Commun.},
    author = {Hollenbach, Michael and Klingner, Nico and Jagtap, Nagesh S. and Bischoff, Lothar and Fowley, Ciarán and Kentsch, Ulrich and Hlawacek, Gregor and Erbe, Artur and Abrosimov, Nikolay V. and Helm, Manfred and Berenc{\'{e}}n, Yonder and Astakhov, Georgy V.},
    number = {},
    month = {12},
    pages = {7683},
    volume = {13},
    publisher = {Nature Research},
    doi = {10.1038/s41467-022-35051-5},
    issn = {20411723},
    pmid = {36509736},
    arxivId = {2204.13173}
}

@article{ChoiZnO2015,
    title = {{Zinc Oxide Nanophotonics}},
    year = {2015},
    journal = {Nanophotonics},
    author = {Choi, Sumin and Aharonovich, Igor},
    number = {4},
    pages = {437--458},
    volume = {4},
    publisher = {Walter de Gruyter GmbH},
    doi = {10.1515/nanoph-2015-0023},
    issn = {21928606}
}
\end{document}